\documentclass[aps,prb,twocolumn,superscriptaddress,amssymb,amstext,showpacs]{revtex4-1}
\usepackage{graphicx,amsfonts,amssymb,amsmath,bm,empheq,color,hyperref,times}

\renewcommand{\vr}{{\bf r}}
\newcommand{\vs}{{\bf s}}

\newcommand{\vm}{{\bf m}}
\newcommand{\vH}{{\bf H}}

\newcommand{\vcH}{{\bf h}}

\newcommand{\half}{\mbox{$\frac{1}{2}$}}

\newcommand{\ez}{{\mathbf{e}_z}}
\newcommand{\erho}{{\mathbf{e}_{\rho}}}
\newcommand{\ephi}{{\mathbf{e}_{\phi}}}

\newcommand{\m}{\mathbf m}

\newcommand\hhalf{{\textstyle\frac12}}
\newcommand\fourth{{\textstyle\frac14}}

\begin{document}

\title{Dzyaloshinskii-Moriya domain walls in magnetic nanotubes}
\author{Arseni Goussev}
\affiliation{Department of Mathematics and Information Sciences, Northumbria University, Newcastle Upon Tyne, NE1 8ST, UK}
\author{J.~M. Robbins}
\author{Valeriy Slastikov}
\affiliation{School of Mathematics, University of Bristol, Bristol, BS8 1TW, UK}
\author{Oleg~A. Tretiakov}
\email{olegt@imr.tohoku.ac.jp}
\affiliation{Institute for Materials Research, Tohoku University, Sendai 980-8577, Japan}
\affiliation{School of Natural Sciences, Far Eastern Federal University, Vladivostok 690950, Russia}

\begin{abstract}
We present an analytic study of domain-wall statics and dynamics in ferromagnetic nanotubes with spin-orbit-induced Dzyaloshinskii-Moriya interaction (DMI).  Even at the level of statics, dramatic effects arise from the interplay of space curvature and DMI: the domains become chirally twisted leading to more compact domain walls. The dynamics of these chiral structures exhibits several interesting features.  Under weak applied currents, they propagate without distortion. The dynamical response is further enriched by the application of an external magnetic field: the domain wall velocity becomes chirality-dependent and can be significantly increased by varying the DMI.  These characteristics allow for enhanced control of domain wall motion in nanotubes with DMI,  increasing their potential as information carriers in future logic and storage devices.
\end{abstract}


\maketitle

\section{Introduction}

In recent years, ferromagnetic nanostructures featuring narrow and stable domain walls (DWs) have been in the spotlight of experimental and theoretical research, with an overarching aim to achieve more compact spintronic logic and memory devices. \cite{Parkin:racetrack08, Allwood01, Allwood02} In particular, numerous efforts have been focused on DWs in ferromagnetic nanowires, \cite{Yamaguchi04, Beach05, KlauiPRL05, HayashiPRL2006, Thomas2006, Meier07, Moriya08, BeachPRL09, Tretiakov08, Rhensius10, IlgazPRL10, Singh10, Krivorotov10, Thomas2010, GRS10Domain, Duine07, Tserkovnyak2008, Lucassen09, Nakatani03, Brataas2010, Tretiakov2012, Linder2014, Hertel2010, Min10, depassier15, benguria_depassier16} nanotubes, \cite{Hertel2011, Hertel2012,  Landeros2012, nanotube_exp2012, nanotube_expPRL2013, nanotube_expPRB2014} and thin films with perpendicular magnetic anisotropy \cite{Thiaville2012, Beach2013, Parkin2013,  Brataas2013, Ohno2014} featuring the Dzyaloshinskii-Moriya interaction  (DMI). \cite{Dzyaloshinskii58, Moriya60} Here we report striking effects arising from the interplay between space-curvature and DMI in ferromagnetic nanostructures, leading to narrow and stable DWs controllable, efficiently and reliably, by means of electric current and magnetic field.

Curvature effects play a significant role in various fields of physics, and are attracting increasing attention in condensed matter, particularly in nanomagnetism. The simplest system with curvature where DW dynamics can be considered is a magnetic nanotube. Furthermore, thin ferromagnetic nanotubes have attracted recent attention from experimentalists owing to a number of technologically advantageous properties, \cite{nanotube_exp2012, nanotube_expPRL2013, nanotube_expPRB2014, Streubel2014, Makarov2014} including enhanced DW stability under strong external fields, allowing for higher DW velocities compared to flat geometries; \cite{Hertel2011, Hertel2012, Hertel2010} increased DW velocities under electric current pulses; \cite{Landeros2012} and the possibility of switching chirality in vortex DWs through magnetic field pulses.\cite{Landeros2012, Landeros_vortex2012}
    
\begin{figure}[ht]
\includegraphics[width=0.99\linewidth]{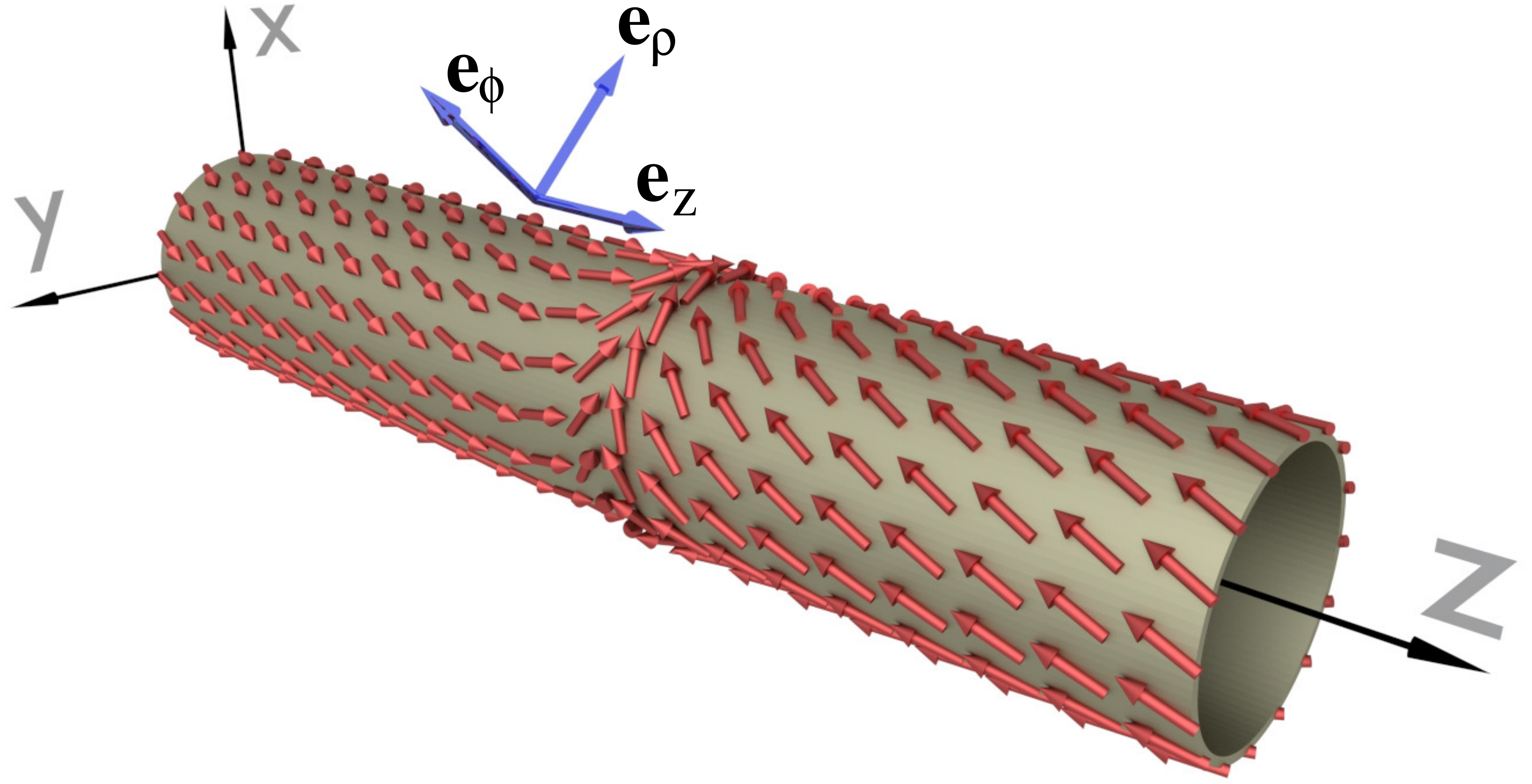}
\caption{(Color online) Domain wall profile in a thin nanotube with Dzyaloshinskii-Moriya interaction. The magnetization lies tangent to the nanotube surface.}\label{fig:DW_profile}
\end{figure}

We show that in thin ferromagnetic nanotubes, the DMI induces qualitatively different effects to those found in flat nanostructures, such as thin films and rectangular nanowires. \cite{Tretiakov_DMI} In nanotubes, DMI causes the domains to become twisted, with helical lines of magnetization as in Fig.~\ref{fig:DW_profile}, forcing the DWs to become narrower. In contrast, in rectangular nanowires with DMI, the magnetization far from the DW remains parallel to the wire axis while DWs 
become broader. This sharpening effect of DMI in nanotubes can enable substantial downscaling in future nanodevices.

We further demonstrate that in a certain thin-nanotube regime specified below, DWs  exhibit perfectly stable motion under  an applied electric current, propagating without any distortion. The adiabatic spin-transfer torque \cite{Tatara04, Zhang04} is 
absent in the spin dynamics equations, and the non-adiabatic term takes the form of the adiabatic one.

Complimentarily to the current, a magnetic field along the nanotube triggers a rich dynamical response in the magnetization texture. We show that the DW velocity becomes strongly dependent on polarity and chirality, \cite{Schmid2013} and can be significantly enhanced by DMI, which is favorable for memory applications. Moreover, the onset of magnonic \cite{Hertel2011} breakdown, impeding DW transport at high fields, can be efficiently suppressed by DMI. 

\section{Statics} 

We consider a ferromagnetic nanotube with inner radius $R$ and thickness $w$. In the thin-nanotube regime $w \ll R$, the micromagnetic energy \cite{Gioia1997} with DMI \cite{Dzyaloshinskii58, Moriya60} takes the form
\begin{multline}
E (\vm) = \int d^3 \vr \, \bigg\{ A |\nabla \vm|^2 +K \left[1 -(\vm \cdot \ez)^2 \right]  \\ 
+DM_s^2 \vm \cdot (\nabla \times \vm) + \frac{\mu_0 M_s^2}{2} (\vm \cdot \erho)^2 \bigg\} \,,
\label{full_energy}
\end{multline}
where the integral runs over the volume of the nanotube, $A$ is the exchange constant, $K$  is the easy-axis crystalline anisotropy, $D$ is the DMI constant, $M_s$ is the saturation magnetization, and $\mu_0$ is  the magnetic permeability of vacuum. The cylindrical-coordinate  unit vectors  $\ez$, $\erho$, and $\ephi$ are shown in Fig.~\ref{fig:DW_profile}.

The last term on the right-hand side of Eq.~(\ref{full_energy}) represents the shape anisotropy stemming from the thinness of the
nanotube. In nanotubes with radius $R$ much larger than the  
magnetostatic exchange length $\sqrt{A/(\mu_0 M_s^2)}$, this term forces the magnetization to lie nearly tangent to the surface.  In this case, the unit vector of magnetization may be described by its orientation $\Theta(z, \rho, \phi; t)$ in the $(z,\phi)$-tangent plane:
\begin{equation}
\label{penalization_of_m}
\vm = \ez \cos\Theta + \ephi \sin\Theta \,.
\end{equation}
Substituting  Eq.~(\ref{penalization_of_m}) into Eq.~(\ref{full_energy}) and introducing the dimensionless coordinates $\vs = \vr/R$, $\zeta = z/R$, $\xi = \rho/R$, anisotropy $\kappa =K R^2 /A$, and DMI constant $\eta = D M_s^2 R / (2 A)$, we thereby obtain the expression for the dimensionless energy $\mathcal{E}=E/(2AR)$:
\begin{equation}
\mathcal{E} = \int d^3 \vs \left( \varepsilon_1(\Theta) + \varepsilon_2(\Theta)\right), 
\label{rescaled_energy}
\end{equation}
where the energy densities $\varepsilon_1$ and $\varepsilon_2$ are given by 
\begin{align}
\varepsilon_1 &= \hhalf \left| \partial_{\zeta}\Theta\right|^2 -V(\Theta) + \fourth (1+\kappa) \,, \label{epsilon-1}
\\ \varepsilon_2 &= \hhalf \left| \partial_\xi \Theta +\eta \right|^2 + \hhalf \left|\partial_\phi \Theta\right|^2\,. 
\label{epsilon-2}
\end{align}
The ``potential'' $V$, which appears in $\varepsilon_1$, is given by
\begin{gather}\label{eq: V}
V(\theta) = \fourth a^2 \cos2 (\theta - \delta),\\ \label{eq: R and delta}
a =\left[(1+\kappa)^2 + 4\eta^2\right]^{\frac14}, \ \ \tan 2\delta = -\frac{2\eta}{1+\kappa},
\end{gather}
where $\delta$ is taken between $-\pi/2$ and $\pi/2$. Below we shall see that $\delta$ determines the orientation of the twisted domains, while $1/a$ is the DW width.

Next we look for a magnetization profile $\Theta$ which minimizes the energy $\mathcal{E}$. The $\varepsilon_2$ term vanishes (and is thus minimized) by taking $\partial_\xi \Theta = -\eta$ and $\partial_\phi \Theta = 0$.  Then  $\Theta$ is of the form
$\theta_0(\zeta) - \eta (\xi -1)$. In the thin-nanotube limit (and taking $\kappa, \eta = O(1)$), the  $\varepsilon_1$ term can be simultaneously minimized by taking $\theta_0(\zeta)$ to satisfy the Euler-Lagrange equation $\theta_0'' = -V'(\theta_0)$, subject to the  boundary conditions that $\theta_0(\pm \infty)$ correspond to maxima of $V$ (not minima). These maxima, given by $\theta = \delta + n\pi$, describe the orientations of magnetization in domains. In the case of zero DMI, ie, $\eta = 0$, the magnetization far from the DW center is parallel to the nanotube axis.   However, for $\eta \neq 0$ the magnetization profile becomes helical, as shown in Fig.~\ref{fig:DW_profile}. 
 
\begin{figure}[ht]
\includegraphics[width=0.99\linewidth]{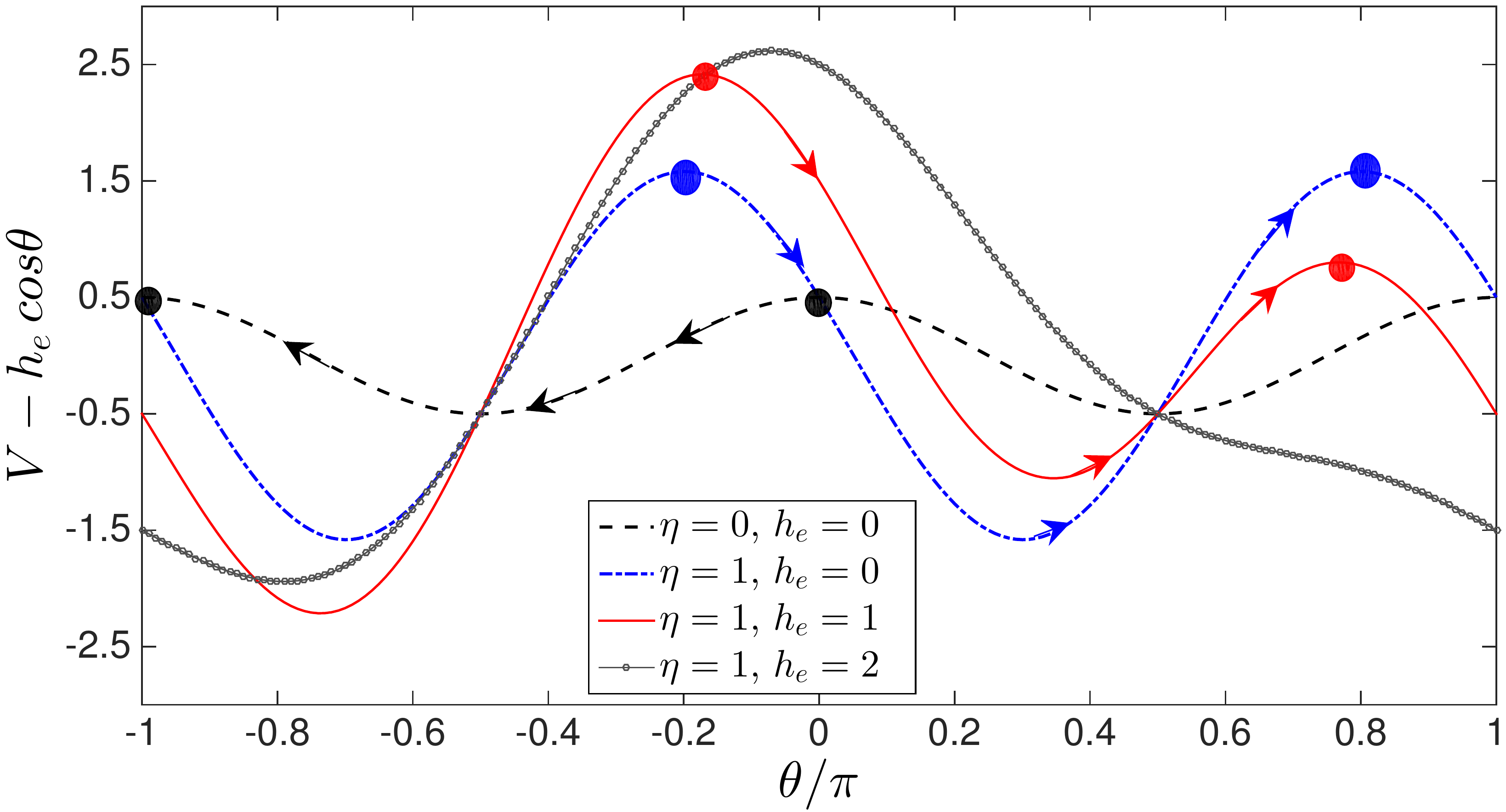}
\caption{(Color online) Mechanical analogy: The profile $\theta(\zeta)$ may be regarded as the instanton orbit of a particle $\theta$ taking infinite time $\eta$ to move between consecutive local maxima of $V(\theta)-h_e\cos\theta$ ($\kappa = 1$ throughout). Dashed curve: With no DMI or applied field, the local maxima are $\theta = 0$ and $\theta = \pi$, corresponding to domains aligned along the nanotube axis. The instanton orbit, indicated by the arrows, describes a tail-to-tail DW with negative chirality. Dotted-dashed curve: With DMI parameter $\eta = 1$ but no applied field, the local maxima are shifted by $\delta = -\pi/8$, corresponding to twisted domains. The instanton orbit corresponds to a head-to-head DW with positive chirality. Solid curve: With $\eta = 1$ and  applied field $h_e = 1$, the values of $V - h_e\cos\theta$ at consecutive maxima are no longer equal; a specific value $j+v$, the coefficient of linear damping in Eq.~\eqref{theta_ODE}, is required to ensure that particle reaches the second maximum without overshooting. Hashed curve: For large enough $h_e$, a maximum and minimum  of $V-h_e\cos\theta$ coalesce, and the instanton orbit is destroyed.}
\label{fig: mechanical analogy}
\end{figure}

Domain walls correspond to boundary conditions $\theta_0(\pm\infty)$ describing oppositely oriented domains. There are four distinct DW profiles, characterized by polarity $\sigma$ and chirality $\chi$ (one is shown in Fig.~\ref{fig:DW_profile}), for more details see Appendix~\ref{appendix1}. Polarity determines whether the DW is head-to-head ($\sigma = 1$) or tail-to-tail ($\sigma = -1$), while chirality determines the sense of rotation of $\vm$ with increasing $\zeta$, so that $\chi = \sigma \operatorname{sgn}  \theta_0'$. The Euler-Lagrange equation may be solved exactly to obtain 
\begin{equation}
\label{eq: static profile}
\theta_0 = 2  \chi \arctan \left( e^{\sigma a \zeta } \right) + \delta .
\end{equation}
The DW profiles may be understood qualitatively in terms of a mechanical analogy -- see Fig.~\ref{fig: mechanical analogy}. We regard $\theta(\zeta)$ as  the trajectory of a  particle  moving in a potential $V(\theta)$ with $\zeta$ playing the role of time. In the static case, the DW boundary conditions correspond to the particle approaching consecutive maxima of $V$ (located at $\delta$ mod $\pi$) as $\zeta \rightarrow \pm\infty$. At times in between, the particle traverses the intervening potential well (this is an example of a so-called instanton orbit).  

An important characteristic of the DW is its width $\Delta$ given by $1/a$, or in physical units, 
\begin{equation}\label{DWwidth}
\Delta = \frac{\sqrt{A}}{\left[ ( K + A/R^2 )^2 + (D M_s^2/R)^2 \right]^{1/4}} \,.
\end{equation}
As is clear from this expression, DWs in thin nanotubes become sharper in the presence of DMI, in marked contrast to the case of rectangular nanowires. \cite{Tretiakov_DMI} DWs also become sharper in nanotubes with higher curvature $1/R$. 

\section{Dynamics} 

Under an applied current, the magnetization dynamics in a ferromagnet far below the Curie temperature is described by the Landau-Lifshitz-Gilbert (LLG) equation: \cite{Gilbert55,Gilbert04, Zhang04,Thiaville05}
\begin{equation}
\label{eq:LLG}
\frac{\partial \m}{\partial t} = \gamma \vH \times \m +\alpha \m \times \frac{\partial \m}{\partial t} -J\frac{\partial \vm}{\partial z} +\beta J \vm \times \frac{\partial \vm}{\partial z} \,,
\end{equation}
where $\vH = -(M_s)^{-1} \delta E / \delta \vm$ is the effective magnetic field, $\gamma$ is the gyromagnetic ratio, $\alpha$ is the Gilbert damping constant, $J$ is the current  along the nanotube in units of velocity, and $\beta$ is the nonadiabatic spin-transfer torque parameter.  In the regime  
\begin{equation}\label{eq: thin-tube regime}
w,  \sqrt{\frac{A}{\mu_0 M_s^2}} \ll R
\end{equation}
 (as considered in the static case) and for currents $J$ satisfying
 \begin{equation}\label{eq: R regime 2}
J \lesssim \frac{\gamma A}{M_s R},
\end{equation}
it can be shown that 
 $\vm$ lies nearly tangent to the nanotube, and  Eq.~(\ref{eq:LLG}) reduces to (see Appendix~\ref{appendix2} for the details of this calculation):
\begin{equation}
\label{eq:LL_tangent_main}
\frac{\partial \vm}{\partial \tau} = -\vm \times (\vm \times
\vcH_{\mathrm{t}}) - j \frac{\partial \vm}{\partial \zeta} \,.
\end{equation}
Here $\tau = \frac{2 \gamma A}{\alpha M_s R^2} t$ is the dimensionless time, $\vcH_{\mathrm{t}}$  is the tangential component of the dimensionless effective field $\vcH = \frac{M_s R^2}{2 A} \vH$, and $j = \frac{M_s R}{2 \gamma A} \beta J$  is the
dimensionless current. Note that $j$ is proportional to nonadiabatic torque parameter $\beta$, whereas the current term itself assumes the adiabatic-torque form. This has an important consequence, as described below.

Proceeding as in Eq.~\eqref{penalization_of_m}, we write $\vm = \ez \cos \Theta + \ephi \sin
\Theta$ with $\Theta(\zeta,\xi,\tau) = \theta(\zeta, \tau) + \eta (\xi -1)$ to obtain
\begin{equation}
\frac{\partial \theta}{\partial \tau} = \frac{\partial^2 \theta}{\partial \zeta^2} - j \frac{\partial \theta}{\partial \zeta} +V'(\theta) \,.
\label{theta_PDE}
\end{equation}
We look for traveling-wave solutions of the form $\theta(\zeta,\tau)=\vartheta(\zeta -v \tau)$ describing axially symmetric DWs propagating with velocity $v$. From Eq.~\eqref{theta_PDE}, the profile $\vartheta$ satisfies
\begin{equation}\label{theta_ODE}
\vartheta''  =(j-v) \vartheta' -V'(\vartheta)
\end{equation}
subject to the same  boundary conditions as in the static case. It is easy to see that the moving profile $\vartheta$ coincides with the static profile $\theta_0$ in Eq.~(\ref{eq: static profile}) with velocity $v = j$. In physical units, the DW velocity is given by
\begin{equation}
\label{eq: V with current}
  \mathcal{V} = \beta J/\alpha \,.
\end{equation}
From Eqs.~\eqref{eq: thin-tube regime} and \eqref{eq: R regime 2}, Eq.~\eqref{eq: V with current} holds for velocities in the regime 
\begin{equation}\label{eq: R regime 4}
{\mathcal V} \ll  \gamma \frac{\beta}{\alpha} \sqrt{\mu_0 A}.  
\end{equation}
Thus, under an applied current, the DW propagates without distortion and with velocity independent of polarity, chirality and DMI. \footnote{See Supplementary Material for the movie of current driven DW.} 
As the velocity approaches $\gamma \frac{\beta}{\alpha} \sqrt{\mu_0 A}$, the magnetization acquires a non-negligible radial component, and new behavior can be expected to appear\cite{Hertel2011, Hertel2012, Landeros2012}.

Next we study the effect of an external magnetic field on DW dynamics. We take the field to be uniform along the nanotube axis,  $\vH_e = H_e \ez$. In the thin-nanotube limit, the applied field generates an additional term in Eq.~(\ref{theta_ODE}):
\begin{equation}
\label{theta_ODE_with_field}
\vartheta'' =(j-v) \vartheta' -V'(\vartheta) - h_e \sin\vartheta \,,
\end{equation}
where $h_e = \frac{M_s R^2}{2 A} H_e$. The boundary conditions are modified so that $\vartheta(\pm\infty)$ correspond to consecutive maxima of a modified potential, $V(\vartheta) -h_e \cos\vartheta$. In terms of the mechanical analogy (Fig.~\ref{fig: mechanical analogy}), $\vartheta(\xi)$ again describes the trajectory of a particle moving from one potential maximum to another, as above. However, the potential difference at consecutive maxima induced by the applied field is compensated now by 
the additional (anti)damping term $(j-v) \vartheta'$.

\begin{figure}[ht]
\includegraphics[width=0.99\linewidth]{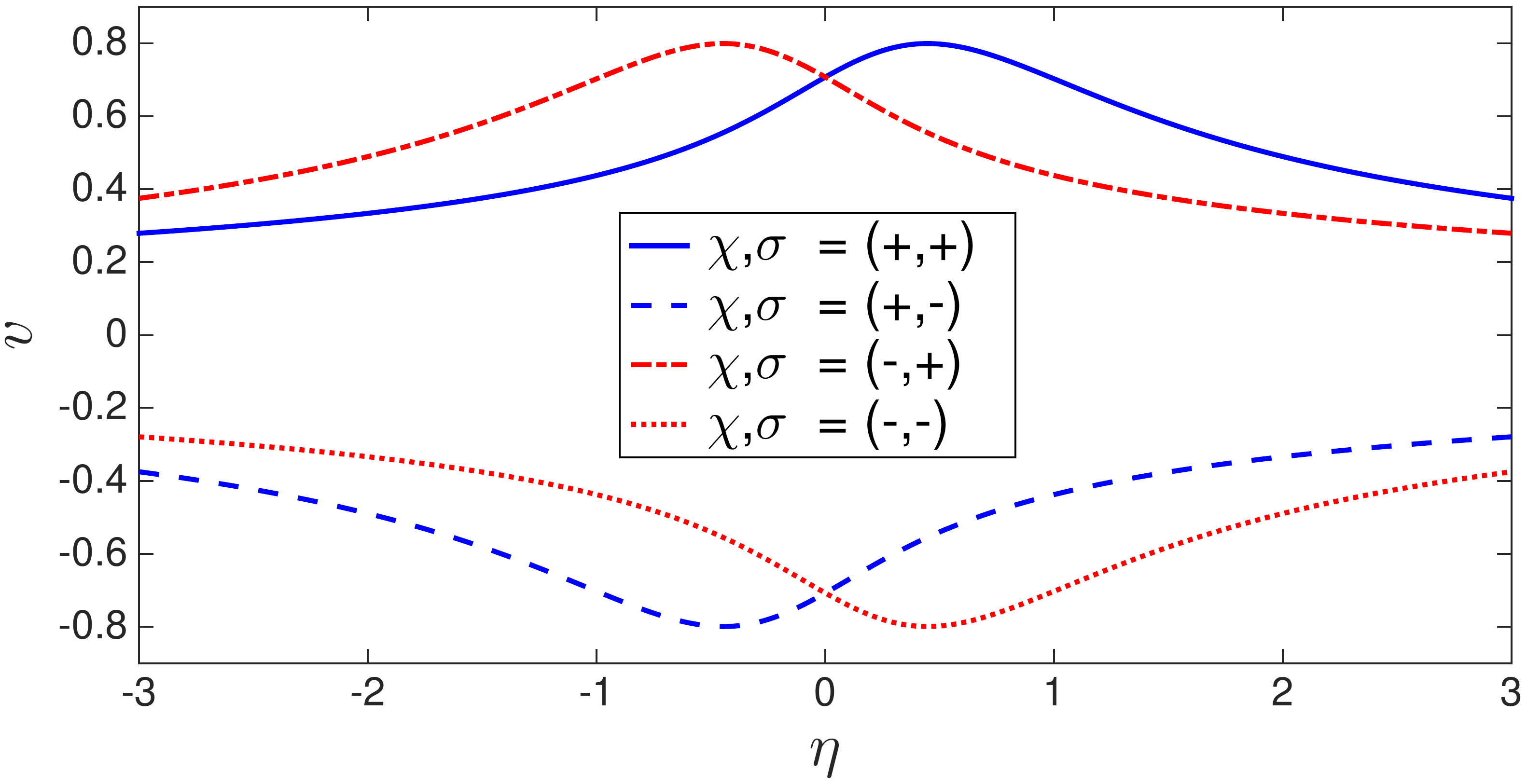}
\caption{(Color online) DW velocity $v$ vs DMI parameter $\eta$ for different chiralities $\chi$ and polarities $\sigma$ ($h_e = \kappa=1$ and $j=0$). Changing the sign of $\chi$ and $\eta$ leaves $v$ unchanged, while changing the sign of $\sigma$ and $\eta$ alters the sign of $v$.} \label{fig: v_vs_eta_numerical}
\end{figure}
\begin{figure}[ht]
\includegraphics[width=0.99\linewidth]{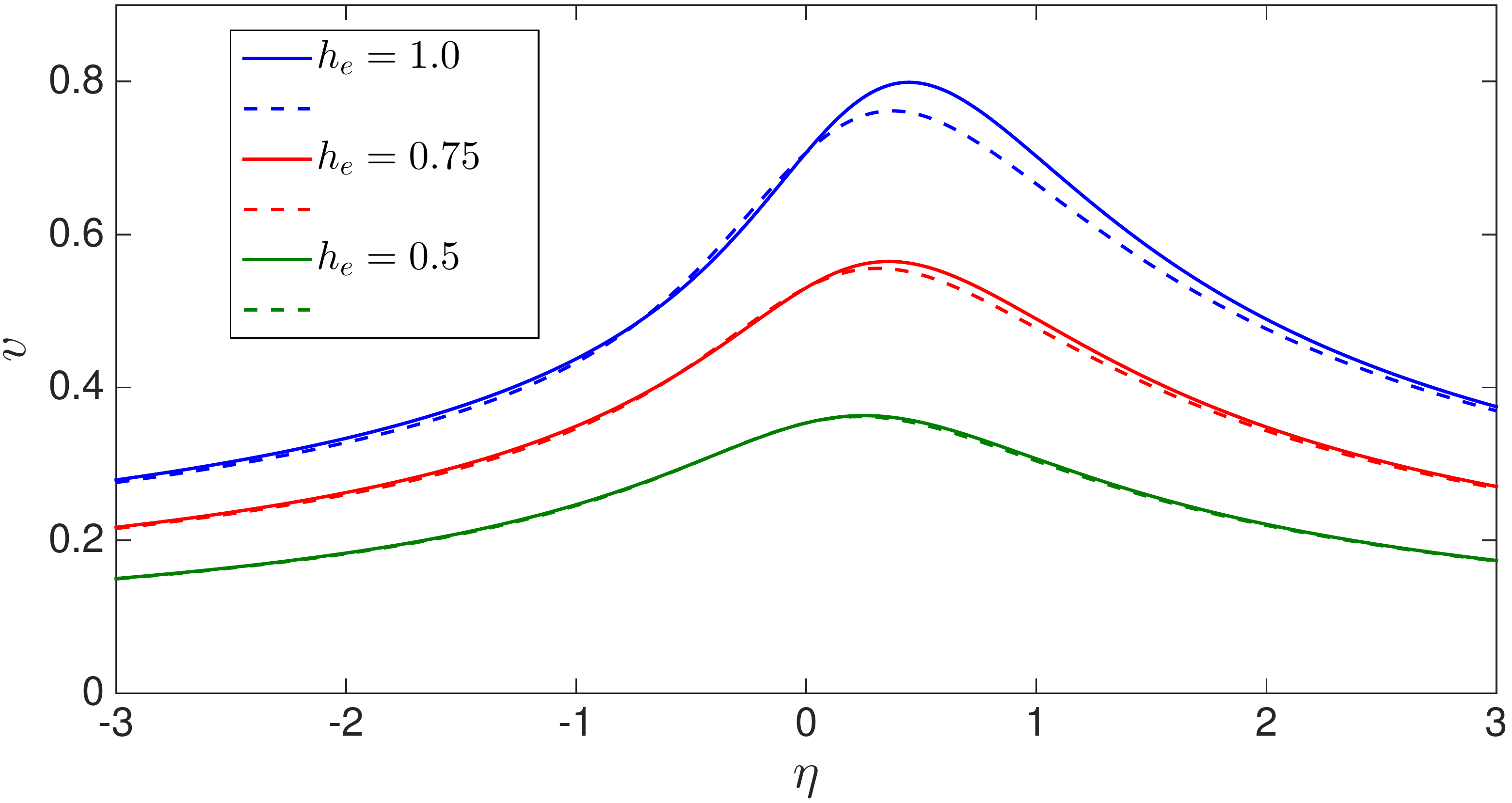}
\caption{(Color online) DW velocity $v$ vs DMI parameter $\eta$ for different values of the external field $h_e$ ($\sigma = +$, $\chi = +$, $\kappa = 1$, and $j=0$ throughout).  The solid curves are obtained from the numerical solution of Eq.~\eqref{theta_ODE_with_field}; the dotted curves are given by the approximate analytical formula~\eqref{eq: v to second order }.}\label{fig: v_vs_eta_analytic}
\end{figure}

Numerical solutions of Eq.~\eqref{theta_ODE_with_field} show that the applied field causes the DW velocity to depend strongly on DMI, chirality, and polarity, see Fig.~\ref{fig: v_vs_eta_numerical}. For a given field strength, the velocity achieves a maximum for a nonzero value of $\eta$, and varies with DMI through $\eta$ by a factor exceeding 2. While Eq.~\eqref{theta_ODE_with_field} cannot be solved analytically, one can develop an expansion in powers of $h_e$ (see Appendix~\ref{appendix3} for the details):
\begin{equation}\label{eq: v to second order }
v = j + \sigma \, \frac{\sqrt{1+\kappa+a^2} }{2a^2} h_e + \chi \frac {\pi\eta}{2a^5}\, h_e^2 .
\end{equation}
As shown in 
Fig.~\ref{fig: v_vs_eta_analytic}, this quadratic approximation is in good agreement with the numerical results.  In the limit of no current and DMI, $j =\eta =0$, it yields $v=\sigma h_e/(\sqrt2 a)$ in accord with Ref.~\onlinecite{Goussev2014}, or in physical units the DW velocity due to magnetic field reads
\begin{equation}
\mathcal{V}=  \frac{\sigma \gamma}{\sqrt{2}\alpha } \frac{ R}{\sqrt{1 + KR^2/A}} H_e .
\end{equation}
In the limit of $R\to\infty$ this expression reduces to a well known result for the velocities of transverse DWs in flat nanostrips. \cite{Tretiakov08}

\begin{figure}[ht]
\includegraphics[width=0.99\linewidth]{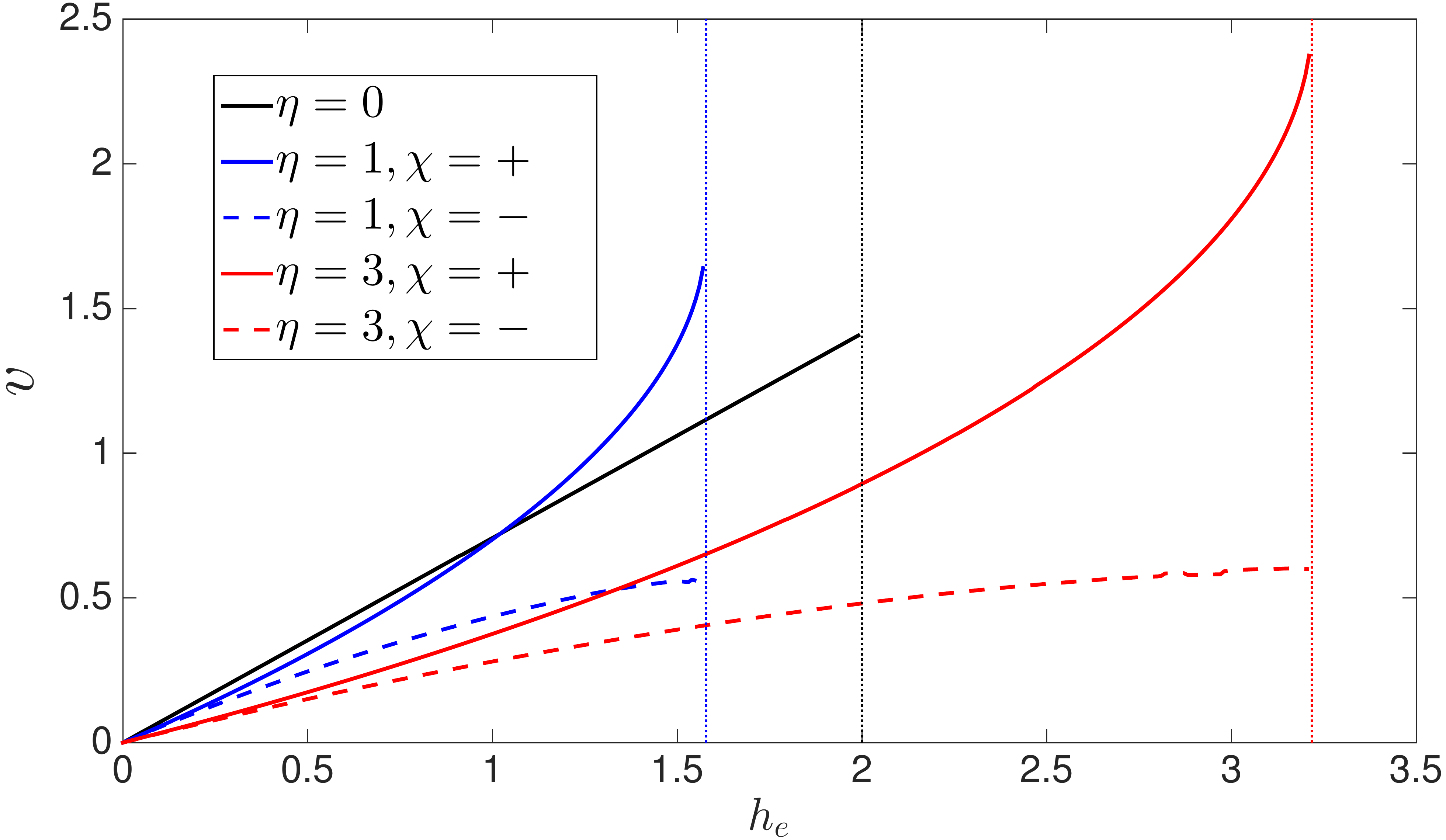}
\caption{(Color online) DW velocity $v$ vs. applied field $h_e$ for different values of the DMI parameter $\eta$ and chirality $\chi$ ($\sigma = +$, $\kappa = 1$, and $j=0$).  For $\eta = 0$, $v$ is given by the exact linear relation $v =  \sigma  h_e/a$.  For $\eta\neq 0$, $v$ is obtained by solving Eq.~\eqref{theta_ODE_with_field} numerically. Curves are computed up to the critical field $h_c$.}
\label{fig: v_vs_h(kappa=1)}
\end{figure}

The dependence of the DW velocity on applied field and chirality is shown in Fig.~\ref{fig: v_vs_h(kappa=1)}. It follows from Eqs.~\eqref{eq: v to second order } and \eqref{eq: R and delta} that for small fields, the DW velocity is suppressed by the DMI.  However, for larger fields and chirality $\chi = \sigma\operatorname{sgn}(\eta h_e)$, the velocity may be enhanced by DMI (for the opposite chirality, $v$ is always reduced). 

\begin{figure}[ht]
\includegraphics[width=0.99\linewidth]{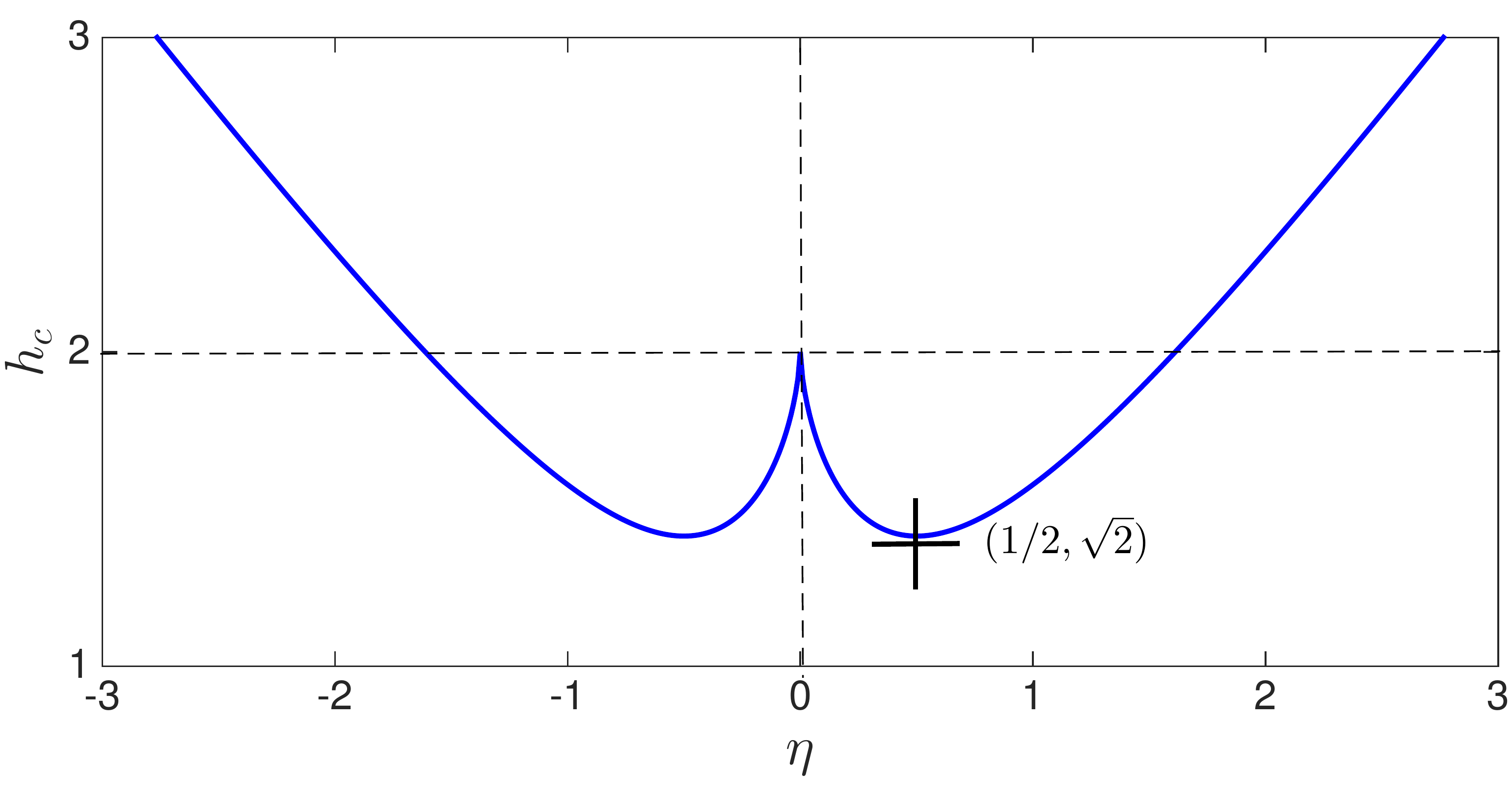}
\caption{(Color online) Critical applied field $h_c$ vs.~DMI parameter $\eta$.} 
\label{fig: hc_vs_eta(kappa=1)}
\end{figure}

At a certain critical applied field $h_c$, a bifurcation occurs, beyond which the DW velocity is suppressed. In terms of  the mechanical analogy of Fig.~\ref{fig: mechanical analogy}, as $h_e$ approaches $h_c$, a maximum and minimum of the potential $V-h_e\cos\vartheta$ coalesce, the instanton orbit is destroyed, and the character of the traveling DW  changes. This phenomenon has been discussed in terms of the spin-Cherenkov effect, \cite{Hertel2011, Landeros2010} and more recently in terms of pulled wavefronts of the KPP equation. \cite{Depassier2014}   It is straightforward to obtain an analytic expression for $h_c$ in terms of $\eta$, shown in Fig.~\ref{fig: hc_vs_eta(kappa=1)}.   For $\eta \ll 1+\kappa$ the leading-order behavior is given by $h_c = 1 + \kappa - 3(c \eta)^{2/3}$, with $c = \sqrt{1+\kappa}/(2\sqrt2)$, while for $\eta \gg 1+ \kappa$ the leading-order behavior is $h_c = \eta$. The important conclusion is that the critical field can be enhanced by increasing DMI, thus allowing for faster DW propagation.

\section{Discussion and Conclusions} 

In recent years there have been ongoing efforts to use ferromagnetic materials with perpendicular anisotropy \cite{Beach2013, Parkin2013,  Ohno2014} to produce sharp and stable domain walls with a view to make potential spintronic logic and memory devices more compact and faster. Here we have described an alternative approach to this goal via domain walls in  thin nanotubes with Dzyaloshinskii-Moriya interaction. These DWs are found to have novel properties: The domains themselves become twisted about the nanotube, forcing the DWs to become sharper with increasing DMI, the opposite of what is seen in thin nanowires. \cite{Tretiakov_DMI} 
 
Under applied currents in the regime specified by Eqs.~\eqref{eq: thin-tube regime} and \eqref{eq: R regime 2}, these DWs propagate without distortion with a velocity proportional to the current.  Applying a magnetic field, we find a rich dependence of  DW velocity on polarity, chirality, DMI and field strength, which may provide enhanced control in future spintronic devices.  The DW velocity can be significantly increased by DMI, and the onset of the magnonic regime suppressed.

This work provides the favorable material trends for engineering nanotubes with DMI for faster and more robust DW operation. Using the DMI parameter from Ref.~\onlinecite{Beach2013} ($DM_s^2=0.5\cdot 10^{-3}$ J/m$^2$), $A=10^{-11}$ J/m, and taking the nanotube radius $R\approx 100$ nm,  we estimate the dimensionless DMI parameter $\eta\approx 2$. For the same material parameters $\kappa=1$ is reached for $K=10^3$ J/m$^3$. These estimates show that the regime where DMI has visible effects is experimentally feasible.
 
In the thin nanotube limit, we are able to treat leading contributions of dipolar interactions exactly. We have derived explicit analytic expressions for the DW profiles and their velocities under applied currents and fields that are in good agreement with numerical solutions of the LLG equation. These results are robust and potentially applicable even beyond the thin-nanotube limit, which is hinted at by recent micromagnetic studies. \cite{Hertel2011}

\acknowledgments

We thank G.E.W. Bauer, J. Barker, G.S.D. Beach, M. Kl\"{a}ui, S.S.P. Parkin, and J. Sinova for helpful discussions. A.G. acknowledges the support of EPSRC Grant EP/K024116/1. J.M.R.~and V.S. acknowledge the support of EPSRC Grant EP/K02390X/1. O.A.T. thanks KITP at University of California, Santa Barbara for hospitality, and acknowledges support by the Grants-in-Aid for Scientific Research (Grants No. 25800184, No. 25247056 and No. 15H01009) from MEXT, Japan; the NSF under Grant No. NSF PHY11-25915; and  SpinNet.

\appendix

\section{Domain walls of different chirality and polarity}\label{appendix1}

\begin{figure*}[ht]
\includegraphics[width=6.5in]{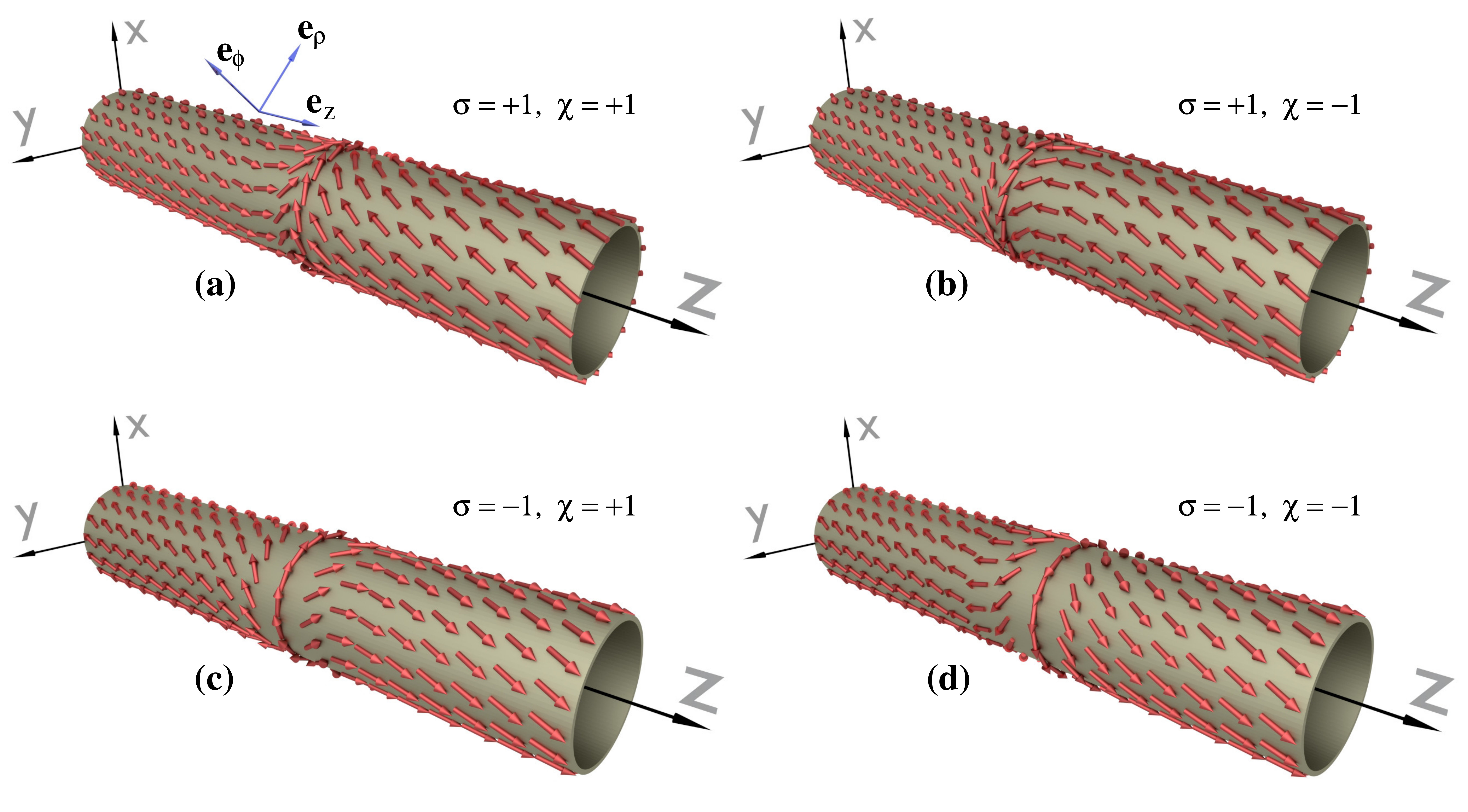}
\caption{(Color online) Domain wall profiles in a thin nanotube with Dzyaloshinskii-Moriya interaction and polarity $\sigma$ and chirality $\chi$.  The magnetization lies tangent to the nanotube surface. }
\label{fig:DW_4profiles}
\end{figure*}

There are four distinct DW profiles, characterized by polarity $\sigma =\pm 1$ and
chirality $\chi =\pm 1$, see Fig.~\ref{fig:DW_4profiles}. The polarity
determines whether the DW is head-to-head ($\sigma = 1$) or
tail-to-tail ($\sigma = -1$), while chirality determines the sense of rotation of $\vm$ with increasing $\zeta$, so that 
$\operatorname{sgn} \theta_0' = \sigma \chi$.

\section{Derivation of Eq.~(\ref{eq:LL_tangent_main}) for current driven domain-wall motion}\label{appendix2}

We start with rewriting the LLG equation, Eq.~(\ref{eq:LLG}), in the Landau-Lifshitz (LL)
form. Taking the vector product of $\vm$ with both sides of the LLG equation, we obtain
\begin{align}
  \vm \times \frac{\partial \m}{\partial t} = &-\gamma 
  \vm
  \times (\vm \times \vH) \nonumber\\ &-\alpha \frac{\partial
    \m}{\partial t} - J \vm \times \frac{\partial \vm}{\partial z} -
  \beta J \frac{\partial \vm}{\partial z} \,.
\label{eq:m_x_LLG}
\end{align}
Then, combining LLG equation and Eq.~(\ref{eq:m_x_LLG}) to
eliminate the $\vm \times \partial \vm / \partial t$ term, we find
\begin{align}
  (1+\alpha^2) \frac{\partial \m}{\partial t} = &-\gamma 
  \vm
  \times \vH - \alpha \gamma 
  \vm \times (\vm \times \vH)
  \nonumber\\ &- (1+\alpha \beta) J \frac{\partial \m}{\partial z} -
  (\alpha - \beta) J \vm \times \frac{\partial \m}{\partial z} \,,
\end{align}
which leads to the LL equation:
\begin{align}
  \frac{\partial \vm}{\partial t} = &-\tilde{\gamma} \vm \times \vH -
  \tilde{\alpha} \vm \times (\vm \times \vH) \nonumber\\ &+ \tilde{J}
  \frac{\partial \vm}{\partial z} + \tilde{\beta} \tilde{J} \vm \times
  \frac{\partial \vm}{\partial z} \label{eq:LL} \,,
\end{align}
where
\begin{align}
  &\tilde{\gamma} = \frac{\gamma 
  }
  {1 + \alpha^2} \,,
  \\ &\tilde{\alpha} = \alpha \tilde{\gamma} = \frac{\alpha \gamma
   }
   {1 + \alpha^2} \,, \\ &\tilde{J} = -\frac{1 + \alpha \beta}{1
    + \alpha^2} J \,, \\ &\tilde{\beta} = \frac{\alpha - \beta}{1 +
    \alpha \beta} \,.
\end{align}

Introducing dimensionless time $\tau'$ through
\begin{equation}
  t = \frac{
  M_s R^2}{2 \tilde{\gamma} A} \tau' = (1+\alpha^2)
  \frac{M_s R^2}{2 \gamma A} \tau' \,,
\end{equation}
the LL equation takes the form
\begin{equation}
  \frac{\partial \vm}{\partial \tau'} = -\vm \times \vcH - \alpha \vm
  \times (\vm \times \vcH) + j_1 \frac{\partial \vm}{\partial \zeta} +
  j_2 \vm \times \frac{\partial \vm}{\partial \zeta} \,,
\label{eq:LL_dimensionless}
\end{equation}
where
\begin{align}
  &j_1 = \frac{
  M_s R}{2 A} \frac{\tilde{J}}{\tilde{\gamma}} = -(1
  + \alpha \beta) \frac{M_s R}{2 \gamma A} J \,, \\ &j_2 = \frac{
    M_s R}{2 A} \frac{\tilde{\beta} \tilde{J}}{\tilde{\gamma}} =
  -(\alpha - \beta) \frac{M_s R}{2 \gamma A} J \,,
\end{align}
and $\vcH = \frac{M_s R^2}{2 A} \vH$, $\zeta = z/R$.

Taking $\epsilon = \frac{2AR^2}{\mu_0 M_s^2} \ll 1$ and  $j_1, j_2 = O(1)$, we have that the radial component of $\vm$ is $O(\epsilon)$, and  
$\vm$ lies
nearly tangent to the nanotube.
Let us decompose
$\vcH$ into its tangential and normal components, $\vcH =
\vcH_{\mathrm{t}} + \vcH_{\mathrm{n}}$, where $\vcH_{\mathrm{t}} =
\vcH - (\vcH \cdot \erho) \erho$ and $\vcH_{\mathrm{n}} = (\vcH \cdot
\erho) \erho$. The LL equation~(\ref{eq:LL_dimensionless}) is now
equivalent to a system of two equations:
\begin{align}
  \frac{\partial \vm}{\partial \tau'} &= -\vm \times \vcH_{\mathrm{n}}
  - \alpha \vm \times (\vm \times \vcH_{\mathrm{t}}) + j_1
  \frac{\partial \vm}{\partial \zeta} \,, \label{eq:LL_t} \\ 0 &= -\vm
  \times \vcH_{\mathrm{t}} - \alpha \vm \times (\vm \times
  \vcH_{\mathrm{n}}) + j_2 \vm \times \frac{\partial \vm}{\partial
    \zeta} \,. \label{eq:LL_n}
\end{align}
Equation~(\ref{eq:LL_t}) is the projection of
Eq.~(\ref{eq:LL_dimensionless}) on the tangent space of the cylinder,
and Eq.~(\ref{eq:LL_n}) is the projection of
Eq.~(\ref{eq:LL_dimensionless}) on $\erho$ direction. From
Eq.~(\ref{eq:LL_n}) we obtain
\begin{equation}
  \alpha \vcH_{\mathrm{n}} = \vm \times \vcH_{\mathrm{t}} - j_2 \vm
  \times \frac{\partial \vm}{\partial \zeta}
\label{eq:h_n-h_t}
\end{equation}
and consequently,
\begin{equation}
  \vm \times \vcH_{\mathrm{n}} = \frac{1}{\alpha} \vm \times (\vm
  \times \vcH_{\mathrm{t}}) + \frac{j_2}{\alpha} \frac{\partial
    \vm}{\partial \zeta} \,.
\label{eq:h_n-h_t-2}
\end{equation}
The substitution of Eq.~(\ref{eq:h_n-h_t-2}) into Eq.~(\ref{eq:LL_t})
yields
\begin{align}
  \frac{\partial \vm}{\partial \tau'} &= -\left( \alpha +
  \frac{1}{\alpha} \right) \vm \times (\vm \times \vcH_{\mathrm{t}}) +
  \left( j_1 - \frac{j_2}{\alpha} \right) \frac{\partial \vm}{\partial
    \zeta} \nonumber \\ &= -\left( \alpha + \frac{1}{\alpha} \right)
  \left[ \vm \times (\vm \times \vcH_{\mathrm{t}}) + j \frac{\partial
      \vm}{\partial \zeta} \right] \,,
\label{eq:LL_tangent}
\end{align}
where
\begin{equation}
  j = -\left( \alpha + \frac{1}{\alpha} \right)^{-1} \left( j_1 -
  \frac{j_2}{\alpha} \right) = \beta \frac{M_s R}{2 \gamma A} J \,.
\end{equation}

Rescaling time once again,
\begin{equation}
  \tau' = \left( \alpha + \frac{1}{\alpha} \right)^{-1} \tau \,,
\end{equation}
so that
\begin{equation}
  t = \alpha \frac{M_s R^2}{2 \gamma A} \tau \,,
\end{equation}
we obtain
\begin{equation}
\frac{\partial \vm}{\partial \tau} = -\vm \times (\vm \times
    \vcH_{\mathrm{t}}) - j \frac{\partial \vm}{\partial \zeta} .
\end{equation}
This is the central equation, Eq.~(\ref{eq:LL_tangent_main}), that we analyze throughout the rest of the paper.

\section{Expansion of domain wall velocity in magnetic field}\label{appendix3}

While Eq.~(\ref{theta_ODE_with_field}) cannot be solved analytically, it is straightforward to develop an expansion in powers of $h_e$.  It turns out that quadratic order is sufficient to capture the leading 
dependence on polarity and chirality.  Letting $p(\vartheta(\zeta)) = \vartheta'(\zeta)$, we may write Eq.~(\ref{theta_ODE_with_field})
equivalently as 
\begin{equation}
 \frac{d}{d\vartheta} \left( \half p^2 + V -h_e\cos\vartheta\right) =  (j-v) p.
\label{eq: first-order equation}
\end{equation}
In terms of the mechanical analogy of Fig.~\ref{fig: mechanical analogy}, 
this corresponds to energy balance. Letting $\epsilon = h_e$, we expand $p = p_0 + \epsilon p_1 + \epsilon^2 p_2$ and $v = v_0 + \epsilon v_1 + \epsilon^2 v_2$.   At zeroth order we deduce that $p_0(\vartheta(\zeta)) = \theta_0'(\zeta)$, where $\theta_0$ is the static profile, i.e. given by Eq.~(\ref{eq: static profile}). Then it follows that $v_0 = j$.  The equations for the next two corrections $p_1$ and $p_2$ can be readily solved, and $v_1$ and $v_2$  are then obtained by integrating Eq.~\eqref{eq: first-order equation} over the interval $\delta < \vartheta < \pi + \delta$ and noting that $p_0$ vanishes at the endpoints. Up to terms of order  $h_e^3/a^5$, we obtain
 \begin{equation}
v = j + \sigma \, \frac{\sqrt{1+\kappa+a^2} }{2a^2} h_e + \chi \frac {\pi\eta}{2a^5}\, h_e^2 .
\end{equation}

\bibliography{micromagnetics_2}
\end{document}